# DYNAMICAL MODEL OF PULSAR-PLANET SYSTEMS

RENU MALHOTRA
*Lunar and Planetary Institute, Houston, TX 77058*

**ABSTRACT**    About two years ago, Wolszczan and Frail announced the detection of a possible planetary system consisting of two Earth-mass planets around a millisecond pulsar. It was pointed out shortly thereafter that the mutual gravitational interaction of the planets leads to predictable evolution of their orbits. Because millisecond pulsars are very stable "clocks" and allow pulse timing measurements with microsecond precision, the orbit evolution would lead to a measurable modulation of the arrival times of the pulses. Detection of this effect would serve to corroborate the planetary interpretation of the data and provide additional constraints on the system. The theory of planetary perturbations in the context of pulsar timing observations is summarized here.

## 1. INTRODUCTION

In an astonishing discovery, Wolszczan & Frail (1992) showed that the quasiperiodic variations of the arrival time residuals of radio pulses from the millisecond pulsar, PSR B1257+12, are most straightforwardly interpreted as the motion of the pulsar about the barycenter of a 'planetary system' consisting of at least two very low mass companions. The authors also showed that a model with two planets on *fixed* Keplerian orbits fitted to the data yields lower limits of about $3M_\oplus$ for the companion masses, and nearly circular orbits (eccentricity $\approx 0.02$) with orbital periods of about 66 days and 98 days (orbital radii less than half an astronomical unit).

It is hardly necessary to point out the significance of the discovery of the first extra-solar planetary system. The nature of this pulsar-planetary system and its relation to planetary systems like our own will undoubtedly be a subject of much debate in the future. However, as a dynamical system it is perhaps indistinguishable from a generic planetary system which one imagines as a dominant central mass with companions much smaller in mass arranged in well-separated, nearly circular and nearly co-planar orbits. As such, the well-developed science of celestial mechanics can be brought to bear on this system to analyze its orbital dynamics.[1] Indeed, soon after the announcement of the discovery, it was pointed out that the presence of more than one planet allowed a test of the 'planetary' interpretation of the observations (Rasio *et al.* 1992a, Malhotra *et al.* 1992). The mutual perturbations of the planets would be reflected in additional systematic variations of the pulse arrival times which, if detected, would serve to corroborate the planetary hypothesis. This theoretical prediction may now be close to realization (see Wolszczan, these proceedings; also Wolszczan 1994).

---

[1] For an introduction to celestial mechanics, the reader is referred to Danby (1988).

A detailed analytic treatment of the mutual perturbation effects on the pulsar motion and the pulse timing residuals is given in Malhotra (1993). (See also Malhotra 1992 and Rasio *et al.* 1992b.) A numerical analysis of the system – numerically integrating the equations of motion of the three-body system for a specific time interval, determining the best-fit *fixed* Keplerian orbital parameters over that time interval, and then determining the differences from the *true* (perturbed) orbits – has been carried out by Peale (1993). In this paper, I will give a review of the nature of the mutual perturbation effects in the planetary system of PSR B1257+12. Similar effects can be expected to be present in any planetary system, and the analysis can be carried over to other such systems that may be detected in the future.

## 2. TWO-PLANET ORBITAL STABILITY

The first "reality check" in any tentative identification of a multiple-planet system in pulsar timing analysis should be to determine whether the inferred parameters — in particular, the masses of the planets and their orbital separation — are compatible with dynamical stability of the system. Although there is no rigorous result for the long term stability of an $N$-body ($N > 2$) planetary system, a useful criterion for nearly circular, coplanar orbits is simply that two adjacent orbits be sufficiently well-separated that the mutual perturbations of the neighboring planets do not lead to instability. An approximate result for the circular restricted three-body problem is based upon an analysis of the leading-order perturbations to the orbital eccentricity and semimajor axis. This is the so-called *resonance overlap criterion* (cf. Chirikov 1979) that estimates the locations and widths of first-order orbital resonances of a planet, and determines the orbital radius at which neighboring resonances cease to overlap (Wisdom 1980). This criterion for stability can be written as follows:

$$\frac{|a - a_p|}{a_p} \gtrsim 1.5 \left(\frac{m_p}{m_\star}\right)^{2/7} \tag{1}$$

where $m_\star$ is the mass of the central star, $m_p$ and $a_p$ are the mass and orbital radius of the planet, and $a$ is the orbital radius of a test particle whose stability is to be determined (see Figure I). The numerical coefficient in Eqn. 1 is from Duncan *et al.* 1989.

To apply this to a planetary system, one can consider each planet in isolation to determine the width of the "instability strip" near its orbit, then demand that any neighboring planet's orbit be outside of this unstable zone.

Another (empirical) formula based upon the results of numerical integrations of general three-body systems is as follows (Graziani & Black 1981):

$$\mu < 0.175 \Delta^3 (2 - \Delta)^{-3/2} \tag{2}$$

where

$$\mu = \frac{m_1 + m_2}{2m_\star}, \qquad \Delta = \frac{a_2 - a_1}{(a_2 + a_1)/2},$$

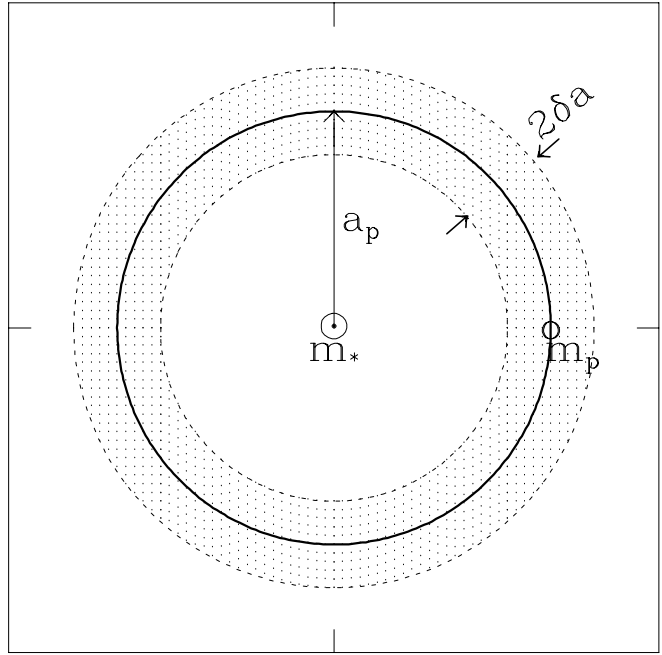

FIGURE I    *The zone of unstable circular orbits near a planet.*

and $m_i, a_i$ are the mass and initial orbital radius of the $i$-th planet.

For the PSR B1257+12 system, the semimajor axes are known within a factor $(\frac{m_\star}{m_{ch}})^{1/3}$ from the observed orbital periods, and the planet masses relative to the pulsar mass are known (from the projected amplitude of the pulsar's motion about the system barycenter) within a factor $[(\frac{m_\star}{m_{ch}})^{1/3} \sin i]^{-1}$, where $i$ is the inclination of the orbital plane to the plane of the sky, and $m_{ch} = 1.4 M_\odot$ (Wolszczan & Frail 1992). Applying either of the above formulas, we can infer that the planetary orbits are stable provided $(m_\star/m_{ch})^{1/3} \sin i \gtrsim 0.003$. Therefore, there exists a strict upper bound on the planet masses, $m/m_\star \lesssim 0.002$ (approximately 2-3 times the mass of Jupiter). An important conclusion follows from this: *the putative pulsar companions are planet-like objects, not low-mass stars.*

## 3. PERTURBATIONS

The planet masses in the PSR B1257+12 system are unlikely to be near the upper limit determined above.[2] Indeed, for $m_\star = m_{ch}$, there is a 95% probability that the two planet masses are no more that 3 times the lower limits of $3.4 M_\oplus$ and $2.8 M_\oplus$. For masses as small as these and the inferred orbits, the orbital perturbations are very small.

---

[2] For random orientation of orbital planes between 0° and 90° to the plane of the sky, the probability that the inclination is less than $i_0$ is $P = 1 - \cos i_0$.

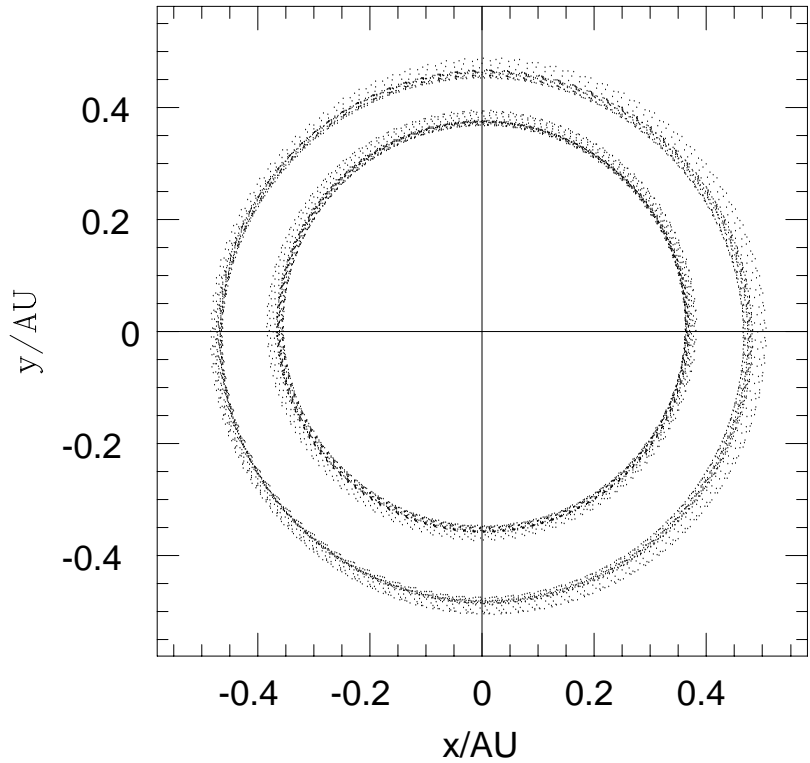

FIGURE II   *The perturbed motion of the planets for a period of 10 years. The position of the planets is plotted at intervals of one day. The magnitude of the perturbations is exaggerated 20-fold for clarity.*

Figures II and III show the results of a numerical integration of the equations of motion for the three-body system with the nominal parameters reported in Wolszczan & Frail (1992). These integrations were carried out using a highly accurate (fifteenth-order) Runge-Kutta-like scheme called *Gauss-Radau* which is especially suitable for planetary integrations (Everhart 1985). An alternative is the Bulirsch-Stoer integrator (see, for example, Press *et al.* 1989) or the symplectic integrators developed recently for Solar System problems (see, for example, Saha & Tremaine 1992 and references therein). These special efforts for numerical solutions are necessary in planetary problems when there is a need for very high accuracy phase information.

Figure II shows the paths traced by the two planets in the PSR B1257+12 system over a period of ten years. The orbital perturbations, $\delta \mathbf{r}(t) = \mathbf{r}(t) - \mathbf{r}_0(t)$ (where $\mathbf{r}(t)$ represents the true motion and $\mathbf{r}_0(t)$ the unperturbed motion), have been exaggerated 20-fold in order to be discernible in this illustration. In Figure III is shown the evolution of the Keplerian orbital elements. There are, in general, three types of dynamical effects apparent in the orbital evolution. These can be identified as (i) short-period effects, (ii) secular effects, and (iii) resonant (more precisely, near-resonant) effects due to the approximate 3:2 ratio of orbital periods. All three contribute comparably to the pulse-arrival time residuals that can be attributed to the mutual interactions of the planets.

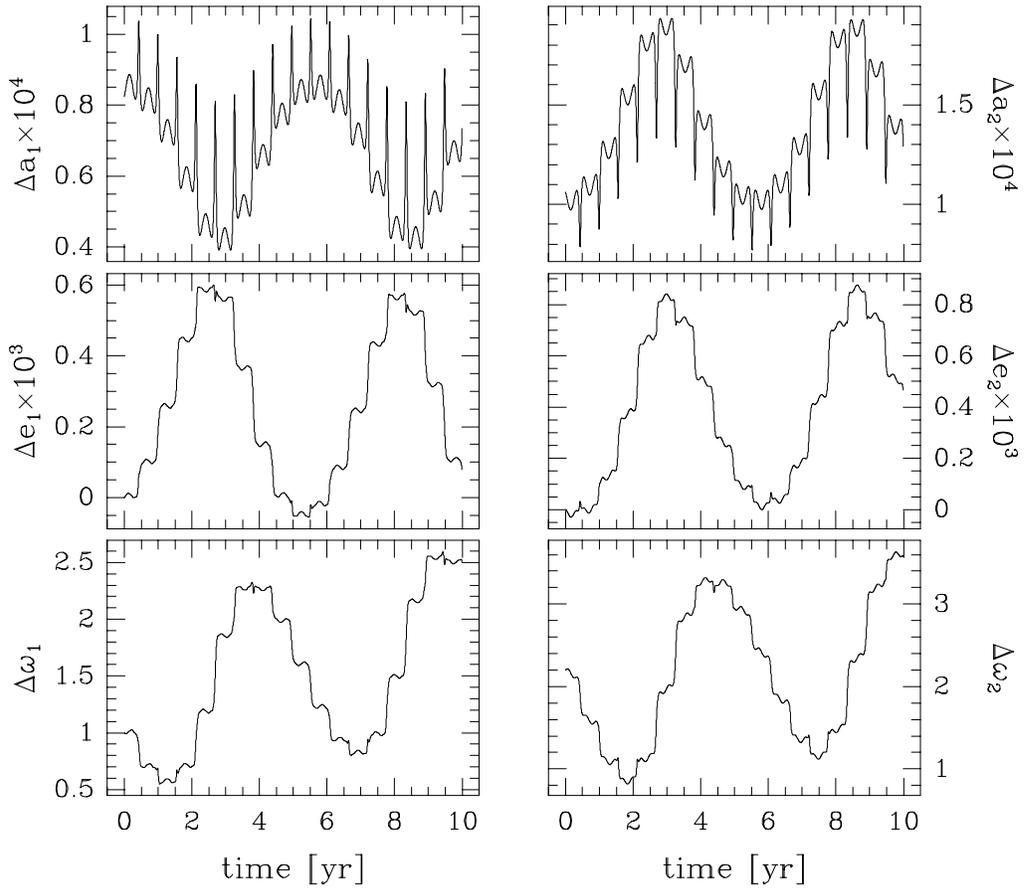

FIGURE III    *The evolution of the Keplerian orbital parameters for a period of 10 years. ($\Delta a_1 \equiv a_1 - 0.3595AU$, $\Delta a_2 \equiv a_2 - 0.4660AU$, $\Delta e_1 \equiv e_1 - 0.022$, $\Delta e_2 \equiv e_2 - 0.02$, $\Delta\omega_1 \equiv \omega_1 - 251$ degrees, $\Delta\omega_2 \equiv \omega_2 - 105$ degrees.)*

*Short-period effects*

The short-period effects are intuitively easy to understand: each time the planets pass each other — at every *conjunction* — they experience a gravitational tug that results in a small change in their orbital velocity, and consequently a small change in their orbital parameters. The frequency of this perturbation is the *synodic* frequency, $n_1 - n_2$, where $n_1$ and $n_2$ are the orbital frequencies of the inner and outer planet, respectively. The spikes in the semimajor axes, and the steps in the eccentricity and periapse (Figure III) are manifestations of the short-period perturbations. We note that the magnitude of the spikes and the steps changes from one conjunction to the next. This is because the distance between randomly oriented ellipses varies with longitude, and, in general, successive conjunctions of the planets occur at different longitudes.

Consider the perturbations of the outer planet on the inner one in the approximation that the outer planet is on a *fixed, circular* orbit, and the inner planet is treated as a massless 'test particle' in an eccentric orbit (in other words, the restricted three-body model). A little reflection shows that the eccentricity variations of the inner planet across conjunctions are as follows: the eccentricity increases at conjunctions that occur when the radial component of the planet's velocity is positive (i.e. from periapse to apoapse), but decreases at conjunctions that occur when the radial component of its velocity is negative (i.e. from apoapse to periapse). The magnitude of the step-like changes in the eccentricity is smallest for conjunctions near periapse and apoapse, and a maximum for conjunctions approximately halfway in-between. Across a conjunction, the net change in the semimajor axis of the inner planet is opposite in phase with respect to its eccentricity changes. The longitude of periapse also suffers step-like changes, but these are out of phase with the eccentricity steps by $\pi/2$. These orbital changes are given by the following approximate expressions:

$$\begin{aligned}
\Delta e &\approx \mathcal{C}(a, a_p) \frac{m_p}{m_\star} \sin M_{\rm c}, \\
\Delta \varpi &\approx -\mathcal{C}(a, a_p) \frac{m_p}{m_\star} \cos M_{\rm c}, \\
\Delta \left(\frac{a_p - a}{a_p}\right)^2 &\approx \frac{4}{3} \Delta e^2, \quad {\rm sign}(\Delta a) = -{\rm sign}(a_p - a){\rm sign}(\Delta e),
\end{aligned} \quad (3)$$

where $a, e$ and $\varpi$ are the semimajor axis, eccentricity and longitude of periapse of the 'test particle', $a_p$ is the semimajor axis of the planet, $M_{\rm c}$ is the mean anomaly (i.e. the mean longitude measured from the periapse) at conjunction, and the coefficient $\mathcal{C}(a, a_p) \sim {\rm sign}(a_p - a)|\frac{a_p - a}{a_p}|^{-\eta}$, $\eta \approx 2$ asymptotically for $\frac{a}{a_p} \to 1$ (cf. Duncan *et al.* 1989).

The short-period perturbations of an outer planet's orbit are given by similar expressions; we note that its semimajor axis perturbation is *in phase* with its eccentricity perturbation.

*Secular effects*

The secular effects are best understood as follows. The *orbit-averaged* gravitational interaction between the two planets corresponds approximately to the interaction potential of two elliptical rings with mass equal to the mass of each planet, and ellipticity and orientation corresponding to the eccentricity and orientation of each Keplerian orbit. (In the simplest approximation, the rings are of uniform density; a more sophisticated approximation would have the rings be of non-uniform mass density to allow for the non-uniform velocity of the planets on Keplerian orbits.) This interaction results in a slow variation of the shape and orientation of the orbits. Mathematically, the leading-order secular perturbations are encapsulated in the variation of the *eccentricity vector*,

$$\mathbf{e}_j = (e_j \cos \varpi_j, e_j \sin \varpi_j) \equiv (k_j, h_j), \quad (4)$$

for each orbit. For small eccentricity and sufficient separation of the orbits, the secular equations for $\dot{\mathbf{e}}_j$ are equivalent to the equations for a set of coupled linear

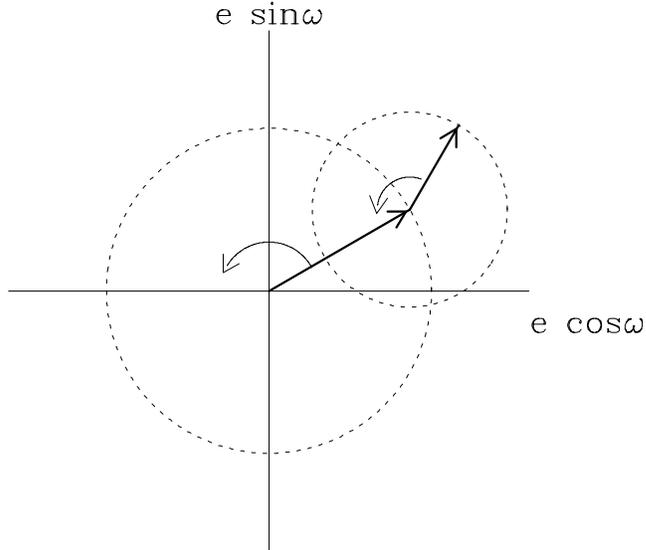

FIGURE IV   *The secular evolution of the eccentricity vector is a linear superposition of two normal modes, represented here as two rotating vectors.*

oscillators:

$$\begin{pmatrix} \dot{h}_1 \\ \dot{h}_2 \end{pmatrix} = \mathbf{A} \cdot \begin{pmatrix} k_1 \\ k_2 \end{pmatrix}, \quad \text{and} \quad \begin{pmatrix} \dot{k}_1 \\ \dot{k}_2 \end{pmatrix} = -\mathbf{A} \cdot \begin{pmatrix} h_1 \\ h_2 \end{pmatrix} \quad (5)$$

where the elements of the two-dimensional array, $\mathbf{A}$, are proportional to the mass ratios $m_j/m_\star$ and the orbital frequencies $n_j$. (The complete expressions may be found in Brouwer & Clemence (1961).) The solution $\mathbf{e}_j(t)$ for a two-planet system is a superposition of two normal-mode oscillations:

$$\begin{aligned} h_i(t) &= \sum_{j=1,2} E_i^{(j)} \sin(g_j t + \beta_j), \\ k_i(t) &= \sum_{j=1,2} E_i^{(j)} \cos(g_j t + \beta_j). \end{aligned} \quad (6)$$

This is illustrated in Figure IV. The two frequencies, $g_1$ and $g_2$, are the eigenvalues of the matrix $\mathbf{A}$. The amplitudes and phases of the two modes are determined by the initial conditions, $(k_j(0), h_j(0))$ at Epoch, inferred from the observations.

A curious property of the PSR B1257+12 planetary system was pointed out by Rasio et al. (1992): the secular solution $\mathbf{e}_j(t)$ for this system shows that approximately 90% of the power is in mode 1. A similar situation is encountered in the Solar System in the secular variations of the Uranian satellites, Titania and Oberon; these satellites are also close to a 3:2 orbital resonance (Dermott & Nicholson 1986, Malhotra et al. 1989). Whether this is merely a coincidence or holds deeper significance (perhaps associated with weak dissipative effects and very long-term evolution in the system) is unknown.

*Resonant effects*

When the ratio of the orbital periods is close to a ratio of small integers that differ by one (i.e. $\frac{j}{j+1}$), the longitude at successive conjunctions changes only a little. This means that the magnitude of the step-like changes in the orbital parameters does not change very much from one conjunction to the next; consequently, the steps accumulate to quite large amplitude variations. The amplitudes of the variations are bounded as the longitude at conjunction must eventually either rotate without bound, or oscillate about some *stable* value. The latter is possible only in a relatively small volume of the phase space. Even so, there are numerous examples of *exact* resonances in the Solar System where the conjunctions of two bodies oscillate about some stable longitude (see, for example, Malhotra 1994). Many of these resonances are believed to arise naturally due to weak dissipative effects that cause slow evolution to stable, phase-locked orbital configurations.

The standard linear perturbation theory breaks down close to a resonance, but a simple picture can still be constructed for the orbital dynamics at a resonance. A brief description of this theory follows. The reader is referred to Henrard & Lemaitre (1983), Peale (1986) or Malhotra (1988) for details.

Sufficiently close to a resonance, it is necessary to take account of the splitting of a resonance due to the fact that, in addition to the two (fast) orbital frequencies, there are two slow degrees of freedom associated with the orbital precession rates. Thus, near a 3:2 commensurability of the orbital periods (or, more generally, any $j+1:j$ commensurability), there are technically at least two distinct resonances defined by the *resonance angles*:

$$\phi_1 = (j+1)\lambda_2 - j\lambda_1 - \varpi_1,$$

$$\phi_2 = (j+1)\lambda_2 - j\lambda_1 - \varpi_2, \tag{7}$$

where $\lambda_i$ are the mean longitudes and $\varpi_i$ are the longitudes of periastron. We define a frequency, $\nu$:

$$\nu = (j+1)n_2^\star - jn_1^\star, \tag{8}$$

where the $n_i^\star$ are auxiliary constants of the system (averaged over the short-period perturbations) and are related to the mean motions and eccentricities as follows:

$$n_1^\star \simeq n_1\left[1 - \frac{3}{2}j\left(e_1^2 + \frac{m_2\sqrt{a_2}}{m_1\sqrt{a_1}}e_2^2\right)\right],$$

$$n_2^\star \simeq n_2\left[1 + \frac{3}{2}(j+1)\left(e_2^2 + \frac{m_1\sqrt{a_1}}{m_2\sqrt{a_2}}e_1^2\right)\right]. \tag{9}$$

The quantity $\nu$ is a measure of the distance from exact resonance; it is the frequency of $\phi_1$ and $\phi_2$ for unperturbed circular orbits[3]. Secular perturbations account for a small splitting of the $\phi_1$- and $\phi_2$-resonance.

---

[3] In the case of resonances amongst satellites of the giant planets in the Solar System, it is generally necessary to include in the definition of $\nu$ the orbital precession rate induced by the planetary oblateness (Malhotra 1988).

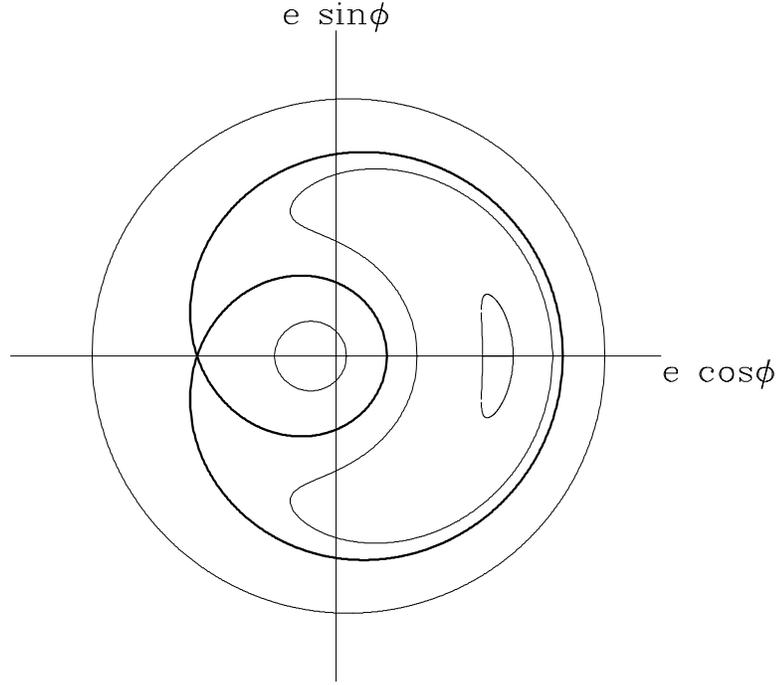

FIGURE V   *The phase space structure near a $j+1:j$ resonance. In the $(e\cos\phi, e\sin\phi)$ phase plane, the trajectories near a resonance are not perfectly circular, but are distorted and offset from the center. The region with banana-shaped trajectories is enclosed by the resonance separatrix (shown as a thick curve) on which the period of motion is unbounded; it defines the region in which the resonance angle $\phi$ may librate.*

Consider each resonance in isolation. Figure V shows the structure of the phase space in a plane on which the polar coordinates are $(e_1, \phi_1)$ or $(e_2, \phi_2 - \pi)$. The bold curve in this figure designates the *resonance separatrix* which separates the region where the resonance angle $\phi_j$ may oscillate (librate) from the regions where it would increase or decrease without bound. (The frequency of $\phi_j$ vanishes on the separatrix.) There are three fixed points in this phase space: (i) a stable fixed point in the *interior zone* near the origin, (ii) a stable fixed point in the *resonance zone*, and (iii) an unstable fixed point which lies on the separatrix.

The phase trajectories in the interior zone near the origin as well those in the zone exterior to the resonance zone are nearly circular but their centers are offset from the origin. Therefore, the eccentricity on these trajectories is not constant but exhibits periodic variations. Away from the resonance separatrix, the *near-resonant* variations of the eccentricity vector are given by

$$\delta \mathbf{e}_i(t) \simeq \frac{\mu_i n_i}{\nu} \Big( \cos\phi_0(t), \sin\phi_0(t) \Big), \tag{10}$$

where $\phi_0(t) = (j+1)\lambda_2 - j\lambda_1$, and $\mu_i$ is a dimensionless parameter that controls the strength of the resonant perturbations:

$$\mu_1 \simeq -\frac{1}{2}\frac{m_2}{m_\star}\alpha\Big[2(j+1) + \alpha\frac{d}{d\alpha}\Big]b_{1/2}^{(j+1)}(\alpha)$$

$$\mu_2 \simeq \frac{1}{2}\frac{m_1}{m_\star}\Big[2j+1 + \alpha\frac{d}{d\alpha}\Big]b_{1/2}^{(j)}(\alpha).$$

(11)

The $b_s^{(j)}(\alpha)$ are Laplace coefficients (Brouwer & Clemence 1961). Similar expressions hold for the perturbations of the semimajor axes.

In the resonance zone, the phase space trajectories are banana-shaped and librate about the stable fixed point whose distance from the origin is $e_{i,res}$ (cf. Eqn. 12 below). The center of stable resonant oscillation is 0 for $\phi_1$, and $\pi$ for $\phi_2$. At exact resonance, there is a relationship between $\nu$, and the resonantly forced eccentricity:

$$\phi_1\text{-resonance}: \quad \nu \approx \frac{3}{2}j^2 n_1 e_{1,res}^2,$$

$$\phi_2\text{-resonance}: \quad \nu \approx \frac{3}{2}(j+1)^2 n_2 e_{2,res}^2.$$

(12)

Note that a solution for $e_{i,res}$ in Eqn. 12 is possible only for $\nu \gtrsim 0$. More precisely, the resonance zone (see Figure V) exists only for $\nu$ greater than a *critical value*, $\nu_{i,c}$, given by

$$\nu_{i,c} \simeq \frac{3}{2}n_i\Big[3j^2\mu_i^2(\alpha,j)\Big]^{1/3}.$$

(13)

The small-amplitude libration frequency is given by:

$$\omega_{i,lib} \approx jn_1\Big|3\mu_i e_{i,res}\Big|^{1/2},$$

(14)

and the resonance half-width for the $\phi_i$-resonance is

$$\Delta\nu_i \approx 2\omega_{i,lib}.$$

(15)

In other words, a resonance libration condition exists provided $\nu$ satisfies one of the conditions in Eqn. 12 within $\pm\Delta\nu_i$.

It may be deduced from the above that the magnitude of the resonant perturbations is largest in the vicinity of the separatrix, where their frequency scales as $|\mu_i e_i|^{1/2}$. The width of the resonance also scales by the same factor.

## 4. DISCUSSION

The inferred orbital parameters and plausible masses of the two putative planets of PSR B1257+12 do not satisfy the *exact-resonance* condition; therefore the standard first-order perturbation theory result of Eqn. 10 for near-resonant perturbations is adequate for this system. The frequency of the resonance angles corresponds to a period of $\sim 5.5$ years, and the amplitude of the near-resonant perturbations is proportional to the mass ratio $m_j/m_\star$. The short-period and secular perturbations also scale similarly.

It is evident from Figure III that the near-resonance condition produces by far the largest orbital perturbations. The complete analytical expressions required for modeling the perturbed orbits are given in Malhotra (1993). The analytical solution allows an assessment of the relative contribution of each of the three types of perturbation on the pulse arrival time residuals. A careful examination shows that all three effects contribute comparably, with the near-resonance effects being slightly dominant.

As discussed in the Introduction, the detection of orbital evolution due to the 'three-body effects' constitutes an observational test of the planetary interpretation with only a few years of systematic timing observations ($\sim 3$ years in the case of PSR B1257+12). Furthermore, it also provides more information about the planetary system, as the discussion below shows.

The signature of the mutual planetary perturbations in the timing observations is sensitive to two parameters, namely $\kappa_1 = m_1/m_\star$ and $\kappa_2 = m_2/m_\star$, *and therefore provides a probe for determining these parameters directly.* In order to fit the *true* (perturbed) orbits to the observations it is necessary to introduce two additional parameters which can be conveniently taken to be $\kappa_1$ and $\kappa_2$. The analytical expressions for the perturbed orbits can then be incorporated into standard timing models. (Although these expressions appear rather cumbersome, they are easy to encode on modern computers.)

Recall that $\kappa_{1,2}$ are the same parameters that determine the amplitude of the pulsar's motion about the system barycenter. However, the pulse timing observations yield only the *projected* amplitude of this motion, $\mathcal{A}_j \equiv \kappa_j a_j \sin i_j$, where $a_j$ are the semimajor axes of the planetary orbits, and $i_j$ the inclination of the orbital planes to the plane of the sky. In practice, the analysis of the observations assuming *fixed* orbits requires the fitting of five parameters for each Keplerian orbit: the amplitude $\mathcal{A}_j$, the mean motion $n_j$, the eccentricity $e_j$, the argument of periapse $\omega_j$, and Epoch of periastron $T_j$ It is easy to see, purely from kinematics, that $\kappa_j$ is then determined up to a factor $(m_\star^{1/3} \sin i_j)$ (cf. section 2 in Malhotra 1993). If the *dynamics* of the system is incorporated into the data analysis (i.e. the data is fitted for perturbed orbits rather than fixed orbits), it would yield the absolute values of $\kappa_1$ and $\kappa_2$, and thus also the values of $(m_\star^{1/3} \sin i_j)$.

It should be pointed out that the perturbation analysis summarized in the previous section (as also the solution for the perturbed orbits given in Malhotra 1993) implicitly assumes coplanar orbits. A mutual inclination of the orbits leads to slightly different perturbation amplitudes and the orbit normals precess slowly about the total orbital angular momentum vector of the system. However, these effects on the pulse arrival times are quite small unless the mutual inclination of the two orbits is large. Therefore, a determination of the values of $(m_\star^{1/3} \sin i_j)$ as outlined above also serves to test the co-planarity asumption.

## ACKNOWLEDGMENTS

This research was done while the author was a Staff Scientist at the Lunar and Planetary Institute which is operated by the Universities Space Research Association under contract no. NASW-4574 with the National Aeronautics and Space Administration. This paper is Lunar and Planetary Institute Contribution no. 832.